\documentstyle[preprint,prl,aps]{revtex}
\newcommand{\tj}{$t$-$J$}
\newcommand{\dxy}{$d_{x^2-y^2}$}

\begin{document}
\draft

\title{Ground State Phases of the Doped 4-Leg \tj\ Ladder}
\author{ Steven R.\ White$^1$ and D.J.\ Scalapino$^2$}
\address{ 
$^1$Department of Physics and Astronomy,
University of California,
Irvine, CA 92697
}
\address{ 
$^2$Department of Physics,
University of California,
Santa Barbara, CA 93106
}
\date{\today}
\maketitle
\begin{abstract}

Using density matrix renormalization group techniques, we have
studied the ground state of the 4-leg \tj\ ladder doped near
half-filling.  Depending upon $J/t$ and the hole doping $x$, three
types of ground state phases are found: (1) a phase containing
$d_{x^2-y^2}$ pairs; (2) a striped CDW domain-wall phase, and (3) a
phase separated regime. A CDW domain-wall consists of fluctuating
hole pairs and this phase has significant $d_{x^2-y^2}$ pair field
correlations.

\end{abstract}


\narrowtext

The observation of spin gaps\cite{kojima,batlogg} in the 2-leg
SrCu$_2$O$_3$ and 4-leg La$_2$Cu$_2$O$_5$ ladder compounds and the
recent report of superconductivity in a hole doped
(La,Sr,Ca)$_{14}$Cu$_{24}$O$_{41}$ compound containing CuO$_2$
chains and 2-leg Cu$_2$O$_3$ ladders\cite{uehara} has brought
renewed interest in the properties of even-leg metal-oxide ladders.
Here, making use of recently developed density matrix
renormalization group (DMRG) techniques\cite{dmrg}, we study the
\tj\ model of a 4-leg ladder for a range of $J/t$ values and 
dopings near half-filling.  For the doped 4-leg \tj\
ladder we find three types of ground state phases: (1) a pair-gas phase
containing \dxy\  pairs; (2) a striped CDW domain wall phase, where
each domain wall consists of four holes; and
(3) a phase separated regime.

The Hamiltonian for the \tj\ model is
\begin{equation}
{\cal H} = -t \sum_{\langle ij\rangle,s} P_G \left( c^\dagger_{i,s}c_{j,s} +
c^\dagger_{j,s}c_{i,s} \right)P_G +
J \sum_{\langle ij\rangle} \left( \vec S_{i} \cdot \vec S_{j} - 
\frac{1}{4}\,n_{i}n_{j}\right),
\end{equation}
where $c^\dagger_{i,s}$ and 
$\vec S_i=c^\dagger_{i,\alpha}\vec\sigma_{\alpha\beta}c_{i,\beta}$ 
are electron creation and spin
operators respectively, $n_{i}$ is the occupation number operator,
$P_G$ is the Gutzwiller projection operator which excludes
configurations with doubly occupied sites, and $\langle ij\rangle$
denotes nearest neighbor sites.  Here we report results for ladders
with open boundary conditions for hole dopings of $0 \le x \le 0.25$
and various $J/t$ values.  Our calculations for the 4-leg ladders
were carried out using a DMRG method in which transformation
matrices were stored and used to construct the initial state for
each superblock diagonalization\cite{cavo}.  Of order $10^3$ states
were kept per block, and the final transformation matrices were used
to calculate the ground state expectation values of the desired
operators at the end of the calculation.

The types of ground state phases which we have found are illustrated
in Fig.~1.  The results shown in Fig. 1 are for a $20\times4$
lattice with from 8 to 16 holes. These figures represent the
most probable configuration of holes in the system, obtained by
maximizing the ground state expectation value of a hole projection
operator
\begin{equation}
P(l_1, l_2, \ldots) = \prod_{i=1} p(l_i),
\end{equation}
where $p(l) = (1-n_{l\uparrow})(1-n_{l\downarrow})$ is the hole
projection operator for the $l^{\rm th}$ lattice site.  The results
shown in Fig.~1 were obtained by maximizing $\langle P(l_1) \rangle$
over $l_1$, then maximizing $\langle P(l_1,l_2) \rangle$ over $l_2$
with fixed $l_1$, etc., until all the holes have been located.
Although this procedure is not guaranteed to give the maximum of
$\langle P(l_1, l_2, \ldots) \rangle$ over all $\{l_i\}$, we have
not observed any cases in which it appears to fail.  The positions
of the holes are shown as the solid circles in Fig.~1.

These pictures of most-likely hole configurations are representative
of the three phases we have found for dopings $0 < x < 0.25$, with
$0.25 < J/t < 3$.  Fig. 1(a) shows a gas of pairs, which occurs at
low doping levels for a wide range of $J/t$, in this case $x = 0.1$,
$J/t=0.35$.  Domain-wall phases occur at somewhat higher doping
levels, also for a wide range of $J/t$.  Fig. 1(b) shows a 
domain-wall phase for $J/t=0.5$, $x = 0.15$, where the most probable hole
configuration has holes along a diagonal.  For smaller values of
$J/t$, the most probable hole configuration consists of a zig-zag
pattern along the two center chains, as shown in Fig. 1(c) for 
$x = 0.15$, $J/t = 0.25$. We will refer to the domain walls of Fig. 1(b)
and 1(c) as ``transverse'' (1,1) and ``longitudinal'' domain walls,
respectively.  Phase separation, as shown in Fig. 1(d),  where the
holes have all moved to either end of the ladder, occurs for $J/t$
greater than about 1.5-1.9, in this case $J/t = 2$. Phase separation
first manifests as an attraction between domain walls, and as an
attraction between the ends of the ladder and the walls, as shown
in the figure. For $J/t \sim 3$, the holes become closely
packed at the ends of the ladder\cite{sticking}.

In order to obtain a clearer picture of the nature of the pair-gas
and domain-wall phases, we have examined various local correlations.
Figure 2(a) shows the probability of various hole configurations
near the most likely configuration for the system shown in Fig.
1(a).  The diameter of the dots is proportional to the probability
of the last hole being on that site, when all the other
hole positions are fixed. In this case the left-hand hole of the second
pair from the left is allowed to vary.  Although the maximum point
shown in Fig. 1(a) has this pair as nearest neighbors, the
probability of the last hole being on either the site above or below 
the maximum point is nearly as large. The results are consistent
with Lanczos calculations for two holes on a periodic 
$\sqrt{26}\times\sqrt{26}$ lattice, in
which for $J/t=0.35$ the holes are about 20$\,$\% more likely to be
found across a diagonal than on near-neighbor
sites\cite{prelov,poilblanc}.  Figure 2(b) shows the expectation
value of the kinetic energy on each bond when the location of all
but one of the holes [the same hole as in (a)] has been specified by
the projection operator.

The expectation value of $\vec S_{i}\cdot\vec S_{j}$ near the paired
holes in the two most likely configurations of Fig. 2(a) is shown in
Fig.~2(c) and (d). In these plots, the width of the lines is proportional
to the bond strength $-\langle \vec S_{i}\cdot\vec S_{j}\rangle$.   In
addition to showing the nearest-neighbor correlations, we show
next-nearest neighbor correlations when both sites are adjacent to
the same hole, but only when these correlations are
antiferromagnetic, $\langle \vec S_{i}\cdot\vec S_{j}\rangle < 0$.
Antiferromagnetic correlations coupling next-nearest neighbor sites
across dynamic holes is an almost universal feature of the doped
$t$-$J$ model\cite{holestructures}, and presumably other doped
antiferromagnets. These frustrating correlations develop in order to
minimize the kinetic energy\cite{holestructures}. The strong
diagonal singlet correlation crossing the hole pair in Fig. 2(d) is
a striking example of this effect. The kinetic energy term strongly
favors a singlet bond connecting these sites since for four of the
eight hops available to the holes in this configuration, this bond
becomes a nearest-neighbor exchange bond.  This diagonal singlet is
characteristic of a $d_{x^2-y^2}$ pair\cite{holestructures}.

A closer view of the domain walls in Figs. 1(b) and (c) is shown in
Fig.~3.  The probability of finding at a given site the fourth hole making
up a transverse domain wall is shown in Fig.~3(a).  This shows that
while the (1,1) direction is favored, the domain wall is fluctuating
strongly. At larger values of $J/t$ (e.g. $J/t \sim 1$) the (0,1)
direction becomes favored.  The expectation value of the exchange
field $\vec S_{i}\cdot\vec S_{j}$ for this wall is plotted in
Fig. 3(b) . A number of its features are similar to those of the
hole pair in Fig.~2(d). In this case, instead of one diagonal
singlet bond, three bonds are apparent. These diagonal singlets
allow the wall to fluctuate strongly, reducing the kinetic energy.
Holes bind in pairs in order to share frustrating
bonds\cite{holestructures}. In a pair, however, there is still
frustration present, since the diagonal singlet represents
antiferromagnetic correlations between sites on the same sublattice.
In a transverse domain wall, however, the undoped spin background is
broken into two unconnected parts by the wall, eliminating the
frustration. Application of a staggered magnetic field to one end of
the system (not shown) shows that the domain walls separate
$\pi$-phase shifted regions with short-range antiferromagnetic spin
correlations.  However,
the kinetic energy of the wall is not as low as that of two isolated
pairs, making the walls unstable at low hole densities for moderate
values of $J/t$.

The kinetic energy favors hole configurations that 1) avoid the
edge sites, since the open boundary conditions act like hard walls,
and 2) avoid nearest-neighbor hole positions, since the holes act
like hard-core objects. However, these types of hole configurations
are generally not favored by the exchange energy, leading to
competition.  At weak to moderate $J/t$ values, the ($\pm1$,1) 
directions are favored for the domain wall largely because these hole
configurations avoid nearest-neighbor hole configurations.  

An example of this competition is seen in the most probable location
of a pair in the pair-gas phase. For $J/t=0.5$, pairs are found
primarily on outer chains in order to form undoped two-leg ladder
structures\cite{holestructures}.  An undoped 2-leg ladder has a spin
gap of order 0.5$\,J$, which is associated with both a rise in the
spin excitations and a lowering of the ``vacuum'' ground-state
energy of
the 2-leg ladder\cite{tsunetsugu}.  Thus an undoped 2-leg ladder is
a low-energy configuration.  For $J/t=0.35$, the tendency of the
holes to avoid the edge sites is slightly stronger, and pairs are
more likely to be found on the two middle chains, as shown in Figs.
1(a) and 2(a).

For smaller values of $J/t$, this tendency of the holes to avoid the
edge sites affects the structure of a domain wall, and a
longitudinal domain wall becomes more likely than a transverse
domain wall.  In Fig. 3(c) we show the exchange field near a
longitudinal domain wall. Again, diagonal singlet correlations are
present. In this case singlets are frustrating only near the ends
of the wall.

So far we have characterized the phases of the \tj\ model using the
most probable hole configurations for typical systems. However,
representing a system by a single hole configuration suggests that
the holes are nearly static, which for small or moderate values of
$J/t$ is very far from the truth.  Although with DMRG we can
calculate $\langle P(l_1, l_2, \ldots)\rangle$ for any given
configuration, the space of configurations is too large to study or
portray directly.  An alternative representation of the system can
be obtained by generating a set of configurations chosen randomly
from the probability distribution $\langle P\rangle$. (Since $P$ is
a projection operator, $\langle P\rangle$ is nonnegative.) To
generate these ``typical'' configurations, we have used a simple
classical Monte Carlo algorithm to wander randomly through hole
configuration space according to the probability distribution
$\langle P\rangle$, which is calculated using DMRG.  This Monte
Carlo calculation is done after the DMRG sweeps have finished, and
after we have found the most probable hole configuration, which is
used as the starting point of the Monte Carlo. At each Monte Carlo
step, a hole is chosen at random, as well as one of the four
directions $(\pm 1,0)$, $(0,\pm 1)$. If the move of that hole one
step in the chosen direction is not possible (e.g. a hop onto a
neighboring hole), the step is rejected.
If the step is possible, it is accepted with Metropolis probability 
$\min(1,\langle P'\rangle/\langle P\rangle)$, where the DMRG transformation 
matrices are used to calculate $\langle P'\rangle$.  This
procedure is fast enough to allow several hundred Monte Carlo sweeps,
which is enough to get a number of typical configurations.

In Fig. 4(a) we show 12 typical configurations for a $14\times4$
system with 8 holes and $J/t=0.5$. The first configuration in the
upper left is the initial, most probable one, showing two transverse
domain walls. Moving downward, successive configurations are
separated by 240 Monte Carlo steps, enough to make them nearly
uncorrelated. We see that in most of the configurations, there are
no recognizable domain walls.  From these configurations (and others
not shown) the holes appear to make up a strongly correlated gas,
made up of clusters of two, four, and sometimes three holes. It is
not obvious from the figure that the wavefunction represented by
these configurations should exhibit the charge density wave (CDW)
structure expected from a set of domain walls. 

In Fig. 4(b) we show the total average hole density per rung
$n_r(l)$ for
the system shown in Fig. 4(a). We see that a strong CDW density
variation is present, as one would expect from the maximum
probability domain-wall pictures: the domain walls take up four
rungs, and are separated by two rungs, which form a low-energy
undoped two-leg ladder.  These CDW
domain-wall structures are subtle correlations built into the ground
state wavefunction, and are difficult to see in a limited number of
hole-configuration snapshots, as in Fig. 4(a). The density
variations are usually commensurate with the lattice, with
pronounced two-rung low-doping regions separating hole-rich domain-wall
regions.  The lattice sizes and dopings shown have been chosen to match
and enhance these commensurate density variations.
It is not clear from the results we have so far whether
there is commensurate long-range CDW order at special fillings (such
as $x=1/6$), or simply power-law decay of CDW correlations.
Also shown in Fig. 4(b) are results for a
$24\times4$ system with $J/t=0.35$ and 6 holes, showing 
CDW correlations. In this case there are three separate pairs
which give rise to these ``$4k_F$'' CDW correlations, as opposed to
the two-pair (4 hole) domain-wall structures of Fig. 1(b). This
behavior in the pair-gas phase is similar to the pairing-CDW
correlations observed in 2-chain ladders\cite{twochain}.

In Fig. 4(c) we show results for the equal-time \dxy\  pair-field
correlation function, 
$D(l)=\langle \Delta_d(i) \Delta_d^\dagger(i+l) \rangle$, where
$\Delta_d(i)$ destroys a nearest-neighbor \dxy\  pair at site
$i$\cite{holestructures}.  The figure shows $D(l=10)$ as a function
of doping $x$, with $i_x$ and $i_x+l_x$ chosen symmetrically about
the center of the lattice, and with $i_y=i_y+l_y=2$.  This quantity
is useful as a measure of the overall strength of the pairing
correlations. The pairing correlations for $J/t=0.5$ initially rise
with doping, reaching a maximum between $x = 0.15$ and $x=0.20$, and
then decrease. Extended $s$-wave pairing correlations (not shown)
are much smaller in magnitude.
For $J/t=0.5$ the magnitude of the correlations
near the maximum is similar to that seen in a two-leg Hubbard
ladder with $U=8t$ (corresponding to $J \sim 4t^2/U = 0.5$)\cite{twochain}. 
For $J/t=0.35$ the peak
is reduced in magnitude and occurs at somewhat reduced doping.
For $J/t=0.25$ the correlations (not shown) are less than $10^{-4}$.
The behavior of $D(l)$ versus $l$ near the maximum (not shown) is
consistent with a power law behavior.  
The results shown bear a strong resemblence
to a plot of $T_c$ versus $x$ for a typical cuprate
superconductor\cite{keimer}.  Remarkably, the pairing
correlations are larger in the domain-wall phase than in the
pair-gas phase. The domain-wall phase appears to exhibit
``supersolid'' behavior, with simultaneous pairing and CDW correlations.
From the hole-configuration snapshots, we see how this can happen:
the domain walls appear as an unbound resonance of hole pairs. There
are also weaker resonances involving three-hole structures. These
resonances are not strong enough to significantly weaken the pairing, 
and, in fact, the increased density of pairs in the
domain-wall phase leads to an increase in the pairing correlations
relative to the more dilute pair-gas phase, as seen in Fig. 4(c).

The domain-wall phase we have found resembles in some respects
the singlet striped phase proposed by Tsunetsugu, et. al.
\cite{tsunetsugu} In addition, various Hartree-Fock calculations
\cite{schulz,poilrice,zaanen,verges}, as well as Gutzwiller
variational Monte Carlo calculations\cite{giamarchi} have found
evidence for the formation of domain walls in the 2D Hubbard model. 
The possibility that a CDW domain-wall phase occurs prior to
phase separation was suggested by Prelovsek and Zotos\cite{prelov}
based on studies of four-hole correlation functions on small \tj\ 
clusters.  Our present calculations show that domain-wall CDW
ground state phases can occur in 4-leg \tj\  ladders. The domain
walls should be thought of as highly-fluctuating resonances 
of pairs.  These CDW domain-wall phases have significant \dxy\  pair
field correlations, which are substantially stronger than in the
low-doping pair-gas phase.

\section*{Acknowledgements}

We would like to thank S.A. Kivelson and T.M. Rice for useful
discussions. SRW acknowledges support from the NSF under 
Grant No. DMR-9509945, and DJS acknowledges support from the
Department of Energy under grand DE-FG03-85ER45197, and from the
Program on Correlated Electrons at the Center for Material Science
at Los Alamos National Laboratory.

\newpage

\begin{figure}
\caption{Maximum likelihood hole configurations obtained by
maximizing the expectation value of 
$P(\ell_1,\ell_2,\ldots)$, Eq.~(2), illustrating the three phases
of the doped 4-leg \tj\  ladder.
(a) A gas of pairs with $J/t=0.35$ and a filling of $x=0.1$. 
(b) ``Transverse'' (1,1) domain walls with $J/t=0.5$ and $x=0.15$. 
(c) ``Longitudinal'' domain walls with $J/t=0.25$ and $x=0.15$.
(d) Phase separation with $J/t=2.0$ and $x=0.2$.
}
\end{figure}

\begin{figure}
\caption{This sequence of plots shows a section of the
lattice for the pair-gas phase, Fig.~1(a), with $J/t=0.35$ and $x=0.1$,
containing the second pair from the left.
(a) The probability of finding the second member of a hole pair when the
first has been projected out at the gray shaded position. All of the
holes in other pairs have also been projected out.
The diameter of the black dots is proportional to the probability of
finding the hole on the corresponding site.
(b) The hopping kinetic energy of one member of a pair when the
other is projected out at the shaded site, shown as the width of the
line connecting nearest neighbor sites, according to the scale
shown.
(c)~The expectation value of $\vec S_{i}\cdot\vec S_{j}$ between various
sites when the holes are nearest neighbors and (d) when they are
next nearest neighbors.
}
\end{figure}

\begin{figure}
\caption{A section of a $20\times4$ lattice showing a domain wall,
with $J/t=0.5$ and $x=0.15$, as in Fig. 1(b).
(a) The probability of finding the fourth hole when the others have been
projected out, and (b) the expectation value  of 
$\vec S_{i}\cdot\vec S_{j}$ when all holes have been projected out
in their most likely configuration.
(c) Same as (b), but for the system shown in Fig. 1(c), with $J/t=0.25$.
}
\end{figure}

\begin{figure}
\caption{(a) Typical hole configurations of a $14\times4$ lattice
with $J/t=0.5$, and 8 holes.
(b) The total average hole density on a rung as a function of the 
rung location. The upper curve is for the system shown in (a). 
The lower curve is for a $24\times4$ system with $J/t=0.35$ and 6 holes.
(c) The equal time \dxy\  pair field correlation function $D(l)$ at a
separation of $l=10$ rungs versus doping $x$, for $20\times4$ and 
$16\times4$ systems and $J/t=0.35$ and $0.5$. The number of holes
in each of the systems shown is a multiple of four.
}
\end{figure}


\begin{references}
\bibitem{kojima} K. Kojima, et. al., \prl {\bf 74}, 2812 (1995).

\bibitem{batlogg} B. Batlogg, et. al., Bull. Am. Phys. Soc. 
{\bf 40}, 327 (1995).

\bibitem{uehara} M. Uehara, T. Nagata, J. Akimitsu, H. Takahashi, N.
Mori, and K. Kinoshita, preprint.

\bibitem{dmrg} S.R. White, \prl {\bf 69}, 2863 (1992),
\prb {\bf 48}, 10345 (1993).

\bibitem{cavo} S.R. White, 1996 preprint, cond-mat/9604129.

\bibitem{sticking} For large values of $J/t$, DMRG can become stuck
in metastable hole-cluster configurations, such as all in a cluster
at one end rather than split into two clusters at the ends. To find the 
lowest energy state, one can initialize the DMRG iterations with the holes 
forced into specific locations, allow convergence, and then choose
the calculation with the lowest final energy.

\bibitem{prelov} P. Prelovsek and X. Zotos,
\prb {\bf 47}, 5984 (1993).

\bibitem{poilblanc} D. Poilblanc, \prb {\bf 49}, 1477 (1994).

\bibitem{holestructures} S. R. White and D.J. Scalapino,
preprint, cond-mat/9605143.

\bibitem{tsunetsugu} H. Tsunetsugu, M. Troyer, and T.M. Rice,
\prb {\bf 51}, 16456 (1995).

\bibitem{twochain} R.M.\ Noack, S.R. White, and D.J.\ Scalapino, 
\prl {\bf 73}, 882 (1994).

\bibitem{keimer} B. Keimer, et. al. \prb {\bf 46}, 14034 (1992).

\bibitem{schulz} H.J. Schulz, Journal de Physique, {\bf 50}, 2833 (1989).

\bibitem{poilrice} D. Poilblanc and T.M. Rice, \prb {\bf 39}, 9749 (1989).

\bibitem{zaanen} J. Zaanen and O. Gunnarsson, \prb {\bf 40}, 7391 (1989).

\bibitem{verges} Verges, J.A., et. al., \prb {\bf 43}, 6099 (1991).

\bibitem{giamarchi} T. Giamarchi and C. Lhuillier J.A., 
\prb {\bf 42}, 10641 (1990).



\end{references}
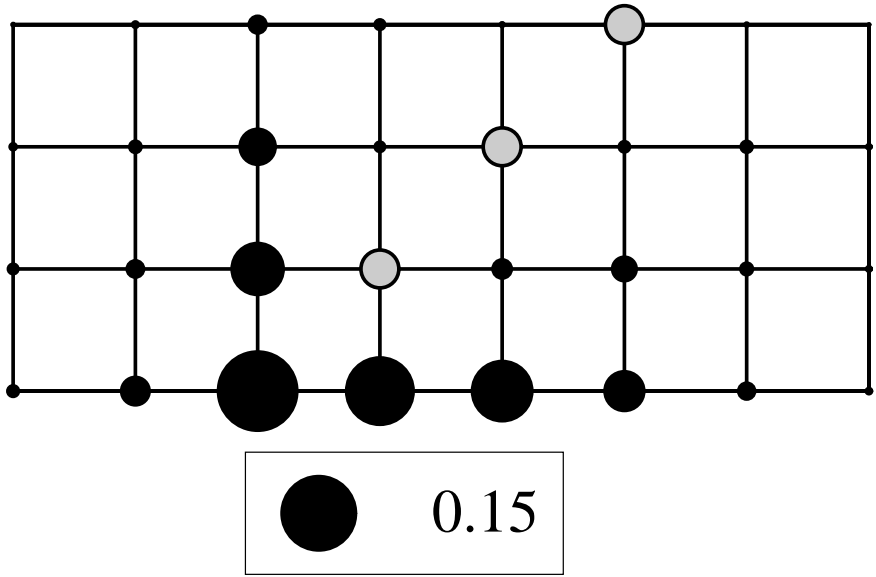
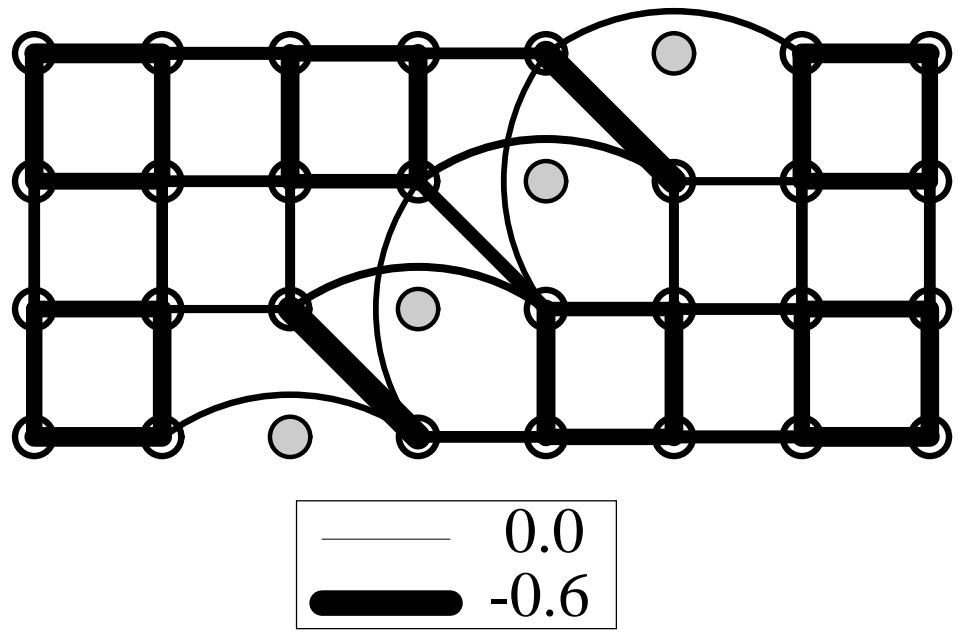
\end{document}